\documentclass[useAMS, referee]{mn2e}
\usepackage{graphicx}
\usepackage{amsmath}

\title[SWIFT J1729.9$-$3437]{ \emph{RXTE} and \emph{Swift} observations of SWIFT J1729.9$-$3437}
\author[\c{S}. \c{S}ahiner, S. \c{C}. \.{I}nam, M. M. Serim, and A. Baykal ]
{\c{S}. \c{S}ahiner$^{1}$\thanks{E-mail: seyda@astroa.physics.metu.edu.tr (\c{S}\c{S}); inam@baskent.edu.tr (S\c{C}\.{I}); 
altan@astroa.physics.metu.edu.tr (AB); muhammed@astroa.physics.metu.edu.tr (MMS)}, S. \c{C}. \.{I}nam$^{2}$\footnotemark[1], 
M. M. Serim$^{1}$ and A. Baykal$^{1}$\footnotemark[1] \\
$^{1}$Physics Department, Middle East Technical University, 06531 Ankara, Turkey\\
$^{2}$Department of Electrical and Electronics Engineering, Ba\c{s}kent University, 06530 Ankara, Turkey}

\begin{document}

\date{Received 2013}

\pagerange{\pageref{firstpage}--\pageref{lastpage}} \pubyear{2013}

\maketitle

\label{firstpage}

\begin{abstract}

We analyse \emph{RXTE} and \emph{Swift} observations of SWIFT J1729.9$-$3437 after its outburst from 2010 July 20 to 2010 
August 12. We calculate a spin frequency and spin frequency derivative of $1.8734(8) \times 10^{-3}$ Hz and 
$6.42(6) \times 10^{-12}$ Hz/s respectively from the quadratic fit of the pulse arrival times.The quadratic fit residuals 
fit well to a circular orbital model with a period of $15.3(2)$ days and a mass function of about $1.3M_{\odot}$, but they 
can also be explained by a torque noise strength of $6.8 \times 10^{-18}$ Hz sec$^{-2}$. Pulse profiles switch from 
double-peaked to single-peaked as the source flux continues to decrease. We find that the pulse shape generally shows no 
strong energy dependence. The hardness ratios reveal that the source becomes softer with decreasing flux. We construct a 
single spectrum from all the available \emph{RXTE} and \emph{Swift} observations. We find that adding an \emph{Fe} line 
complex feature around 6.51 keV slightly improves the spectral fit. We also find that \emph{Fe} line flux correlates with 
X-ray flux which might indicate the origin of the \emph{Fe} emission as the source itself rather than the Galactic ridge. 

\end{abstract}

\begin{keywords}
accretion, accretion discs - stars:neutron - pulsars:individual: SWIFT J1729.9-3437
\end{keywords}

\section{Introduction}

SWIFT J1729.9$-$3437 is a transient X-ray pulsar discovered during all-sky monitoring with the \emph{Swift} Burst Alert 
Telescope (BAT) on 2010 July 13 (Markwardt et al. 2010a). At the same time, \emph{Rossi X-ray Timing Explorer} (\emph{RXTE}) 
Proportional Counter Array (PCA) monitoring scans of the Galactic centre region confirmed a gradual increase in flux from a 
position consistent with the \emph{Swift} position of the source (RA = 262.53746, Dec. = $-$34.61239). Consecutive 
\emph{RXTE}$-$PCA pointings identified SWIFT J1729.9$-$3437 as a pulsar with $\sim$530 s pulsations (Markwardt et al. 2010b). 
Markwardt et al.(2010b) also suggested that the X-ray spectrum of the source is compatible with basic X-ray pulsar spectra, 
modelled by an absorbed power law with a high energy cut-off. 

In this paper, we analyse \emph{RXTE} and \emph{Swift} observations of SWIFT J1729.9$-$3437 and examine the spectral and 
timing properties of its outburst from 2010 July 20 to 2010 August 12 (MJD 55397 $-$ MJD 55420). In Section 2, we describe 
the observations that we analyse. In Section 3 and 4, we present our timing and spectral analysis results. In Section 5, we 
discuss our results and conclude.

\section{Observations}

\subsection{\emph{RXTE}}

11 pointing \emph{RXTE}$-$PCA observations of SWIFT J1729.9$-$3437 were performed between 2010 July 20 and 2010 August 8. 
The total exposure time is about 42 ks. PCA consisted of five co-aligned identical proportional counter units (PCUs) 
(Jahoda et al. 1996) each having an effective area of approximately 1300 $cm^{2}$. The energy range of PCA was from 
2 to 60 keV, with an energy resolution of 18 per cent at 6 keV. The field of view (FOV) at full width at half-maximum (FWHM) 
was about 1$\degr$. Generally, various PCUs were turned off during PCA observations. The responses of PCU0 and PCU1 were not 
well known due to loss of their propane layers. Furthermore, the responses of the top anode layers were modeled better than 
the other layers. Although the number of active PCUs during the observations of SWIFT J1729.9$-$3437 varied between one and 
three, we select only data from the top anode layer of PCU2 because of the reasons mentioned above.

The standard software tools of \verb"HEASOFT v.6.11" are used for the PCA data analysis. Data is filtered by selecting times 
when the elevation angle is greater than 10$\degr$, the offset from the source is less than 0.02$\degr$ and the electron 
contamination of PCU2 is less than 0.1. The background extraction for spectra and light curves are done by using the latest 
PCA background estimator models supplied by the RXTE Guest Observer Facility (GOF), \verb"EPOCH 5C". During the extraction 
of spectra, Standard2f mode data is considered with 128 energy channels and 16-s time resolution. Relevant response matrices 
for spectra are created by \verb"PCARSP V.11.7.1". Furthermore we construct pulse phase resolved spectra with the tool 
\verb"FASEBIN Rev.1.8" by using Good Xenon mode event data and 125 $\mu$s time resolution event files 
(\verb"E_125us_64M_0_1s") when Good Xenon mode data is not available. From these spectra, we generate energy resolved pulse 
profiles by obtaining the count rates per phase bin with the tool \verb"FBSSUM". The 1 s binned 3-20 keV light curves are 
produced from Good Xenon and \verb"E_125us_64M_0_1s" events. 

\subsection{\emph{SWIFT}}

After the discovery of SWIFT J1729.9$-$3437 by the \emph{Swift}$-$BAT (Barthelmy et al. 2005) all-sky monitoring on 2010 
July 13 (Markwardt et al. 2010a), a total of 11 follow-up X-ray Telescope (XRT; Burrows et al. 2005) observations 
(each $\sim$3 ks) were carried out between 2010 July 20 and 2010 August 12. XRT is a focusing instrument on board the 
\emph{Swift} satellite (Gehrels et al. 2004) which operates between 0.2 and 10 keV energy range. XRT has an effective area 
of 110 $cm^{2}$ at 1.5 keV with a spatial resolution of 18$\arcsec$ and the FOV of $23.6\arcmin \times 23.6\arcmin$. 
Operation mode of XRT switches between photon-counting (PC), imaging and timing depending on the brightness of the observed 
source. Pointing observations of SWIFT J1729.9$-$3437 are in the PC mode. Screened event files are produced by the 
\verb"XRTDAS v.2.7.0" standard data pipeline package \verb"XRTPIPELINE v.0.12.6". Standard grade filtering (0-12) is applied 
to PC mode data.

The spectral extraction is carried out by filtering appropriate regions using \verb"XSELECT v.2.4b". For the observations 
in which the XRT count rate is high enough (above 0.5 cts/s) to produce event pile-up, source region is selected to be 
annular. We compared the observed and nominal point spread function (PSF) (Vaughan et al. 2006) and decided the size of 
the core affected in order to determine the radius of the inner circle of the annulus. For this purpose, we first modelled 
the outer wings ($>15 \arcsec$) of the PSF with a King function which has typical parameter values for XRT 
(Moretti et al. 2005). Then the model is extrapolated to the inner region for comparison with the observed PSF and the size 
of the piled-up region is determined from the deviation point between the data and the model. For the brightest observations 
an exclusion region of radius $\sim9\arcsec$ is sufficient to correct pile-up. For low count rate observations a circular 
region of radius 47$\arcsec$ is selected for the source spectral extraction. Source regions are centred on the position 
determined with \verb"XRTCENTROID v.0.2.9". A circular source-free region with 140$\arcsec$ radius is selected for the 
background spectral extraction. Resulting spectral files are grouped to require at least 50 counts per bin using the ftool 
\verb"GRPPHA v.3.0.1", for the $\chi^{2}$ statistics to be applicable. We used the latest response matrix file (version v013) 
and created individual ancillary response files using the tool \verb"XRTMKARF v.0.5.9" with the exposure map produced by 
\verb"XRTEXPOMAP v.0.2.7". Spectral analysis is performed using \verb"XSPEC v.12.7.0".

For the timing analysis, arrival times of XRT events are first corrected to the Solar system barycenter by using the tool 
\verb"BARYCORR v.1.11". Then, background-subtracted XRT light curves in the 0.2$-$10 keV energy range; corrected for pile-up, 
PSF losses and vignetting; are extracted by using the highest possible time resolution (2.51 s) for PC mode data.

\section{Timing Analysis}

\subsection{Timing Solution}

\begin{figure}
  \center{\includegraphics[height=8.4cm, angle=270]{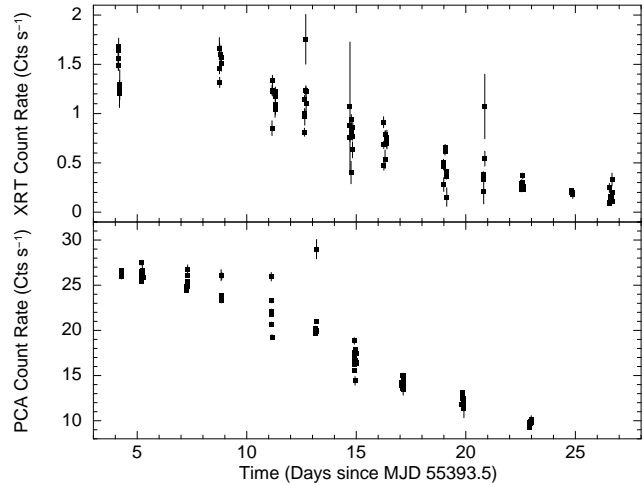}} 
  \caption{530 s binned light curves from \emph{Swift}$-$XRT (0.3$-$10 keV; upper panel) and \emph{RXTE}$-$PCA (3$-$20 keV, 
  PCU2 top layer; lower panel) observations.}
  \label{bothlc}
\end{figure}

For the timing analysis, we use 1 s binned \emph{RXTE}$-$PCA and 2.51 s binned \emph{Swift}$-$XRT light curves of the source. 
To illustrate the temporal variation of the pulse phase averaged count rate of the source, in Fig. \ref{bothlc}, we present 
530 s binned versions of these light curves. These background subtracted light curves are also corrected to the barycenter 
of the Solar system. 

In order to estimate pulse periods of the source, the \emph{RXTE}$-$PCA time series is folded on statistically independent 
trial periods (Leahy et al. 1983). Template pulse profiles are formed from these observations by folding the data on the 
period giving maximum $\chi^2$. Then the barycentred \emph{Swift}$-$XRT time series are also folded over the best period 
obtained from \emph{RXTE}. Pulse profiles consisting of 20 phase bins are represented by their Fourier harmonics 
(Deeter \& Boynton 1985). Using cross-correlation of pulses between template and in each observation, we obtain the pulse 
arrival times.

\begin{figure}
  \center{\includegraphics[width=8.4cm]{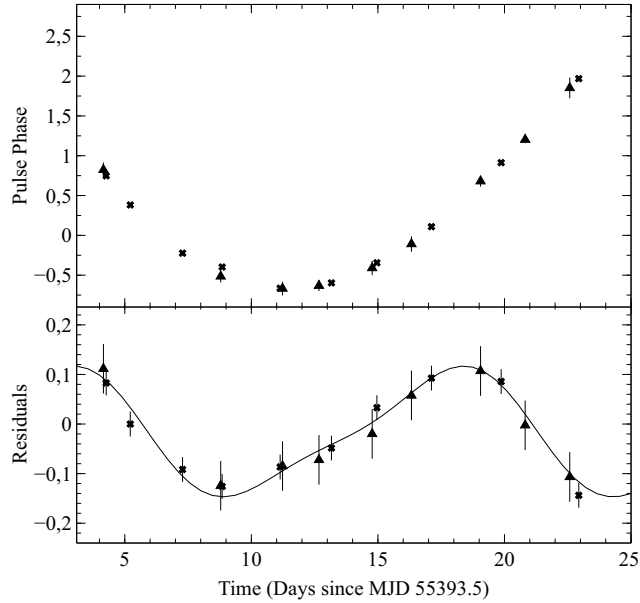}} \vspace*{0.5cm}
  \caption{Pulse phases and their residuals after a quadratic fit. Crosses and triangles represent data from 
  \emph{RXTE}$-$PCA and \emph{Swift}$-$XRT respectively. The solid line in the bottom panel corresponds to the elliptical 
  orbital model with an upper limit of eccentricity of 0.60 and $T_{\pi/2}=5^{\circ}$ and circular orbital parameters 
  listed in Table \ref{soln}.}
  \label{arrt}
\end{figure}

We are able to connect all the pulse arrival times of the observations in phase over the whole observation time span. 
Following the approach of Deeter et al. (1981), we find that it is possible to fit the pulse arrival times to the second 
order polynomial,
\begin{equation}
\delta \phi = \delta \phi_{o} + \delta \nu (t-t_{o}) 
+ \frac{1}{2}\dot{\nu} (t-t_{o})^{2} 
\label{polyn}
\end{equation}
where $\delta \phi$ is the pulse phase offset found from the pulse timing analysis, $t_{o}$ is the epoch for folding; 
$\delta \phi_{o}$ is the residual phase offset at t$_{o}$; $\delta \nu$ is the correction to the pulse frequency at time 
$t_0$; $\dot{\nu}$, being the pulse frequency derivative, is the second derivative of the pulse phase. In Fig. \ref{arrt}, 
we present the pulse phases and the residuals of this quadratic fit. 

\begin{figure}
  \center{\includegraphics[width=8.4cm]{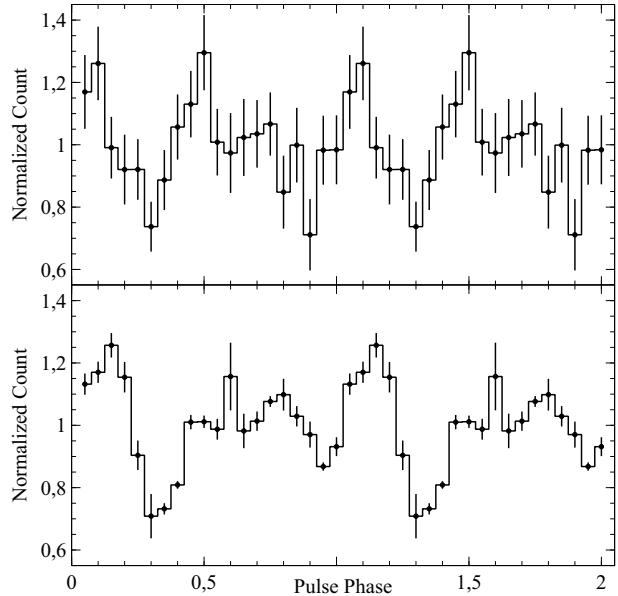}} \vspace*{0.5cm}
  \caption{Two pulse examples obtained from the \emph{Swift}$-$XRT observation at MJD 55397.66 (top panel) and the 
  \emph{RXTE}$-$PCA observation at MJD 55397.77 (bottom panel).}
  \label{pulexp}
\end{figure}

In Fig. \ref{pulexp}, we show typical \emph{Swift}$-$XRT and \emph{RXTE}$-$PCA pulse profiles. As seen from this figure, the 
\emph{Swift} pulse has larger error bars and less signal to noise ratio relative to the \emph{RXTE} pulse. In the pulse 
timing analysis we find that the error bars of the pulse phases of \emph{RXTE} and \emph{Swift} are inversely correlated 
with the signal to noise ratio of the pulses of each observation.

\begin{table}
\caption{Timing Solution and Orbital Model of SWIFT J1729.9$-$3437.}
\center{\renewcommand{\arraystretch}{1.2}\begin{tabular}{|l|l|}
\hline \hline
{\bf{Timing Solution Parameter}} & {\bf{Value}} \\
\hline
Timing Epoch (MJD) & $55393.50(6)$ \\
Frequency (Hz) & $1.8734(8) \times 10^{-3}$ \\
Frequency Derivative (Hz.s$^{-1}$) & $6.42(6) \times 10^{-12}$ \\
\hline \hline
{\bf{Circular Orbital Model Parameter}} & {\bf{Value}} \\
\hline
Orbital Epoch (MJD) & $55395.7(6)$ \\
${{a} \over {c}}\sin i$ (lt.s) & $65(3)$ \\
Orbital Period (days) & $15.3(2)$ \\
\hline
\end{tabular}}
\label{soln}
\end{table}

The residuals shown in Fig. \ref{arrt} fit well to a sinusoidal function with a period of 15.3(2) days. This corresponds to 
a circular orbital model, parameters of which are listed in Table \ref{soln}. This circular model has a reduced $\chi^2$ of 
1.0. Using an elliptical model we find an upper limit for the eccentricity as 0.6 (see bottom panel of Fig. \ref{arrt}). 
In this case, reduced $\chi^2$ is found to be 0.4. This small reduced $\chi^2$ value indicates that the elliptical orbital 
model "over-fits" data.

The residuals of the quadratic can alternatively express the noise process due to random torque fluctuations 
(Bildsten et al. 1997, Baykal et al. 2007). In order to estimate the noise strength, we fit a cubic polynomial to the 
residuals of the pulse arrival times. The observed time series is simulated by the Monte Carlo technique for a unit white 
noise strength defined as $P_{\dot \nu}({f})=1$ and fitted to a cubic polynomial in time. Then the square of the third order 
polynomial term is divided into the value from Monte Carlo simulations (Deeter 1984, Cordes 1980). The torque noise strength 
is obtained as $6.8 \times 10^{-18}$ Hz sec$^{-2}$. This value of noise strength estimate is comparable with those of other 
accretion powered sources such as wind accretors e.g. Vela X$-$1, 4U 1538$-$52 and GX 301$-$2 which has the values changing 
between $10^{-20}$ and $10^{-18}$ Hz s$^{-2}$ (Bildsten et al. 1997). Her X$-$1 and 4U 1626$-$67, which are disc accretors with 
low mass companions, have shown pulse frequency derivatives consistent with noise strengths 
$10^{-21}$ to $10^{-18}$ Hz sec$^{-2}$ (Bildsten et al. 1997). Therefore residuals of quadratic fit can also be associated 
with torque noise fluctuation of the source.

\subsection{Pulse Profiles and Hardness Ratios}

\begin{figure}
  \center{\includegraphics[width=7.7cm, angle=270]{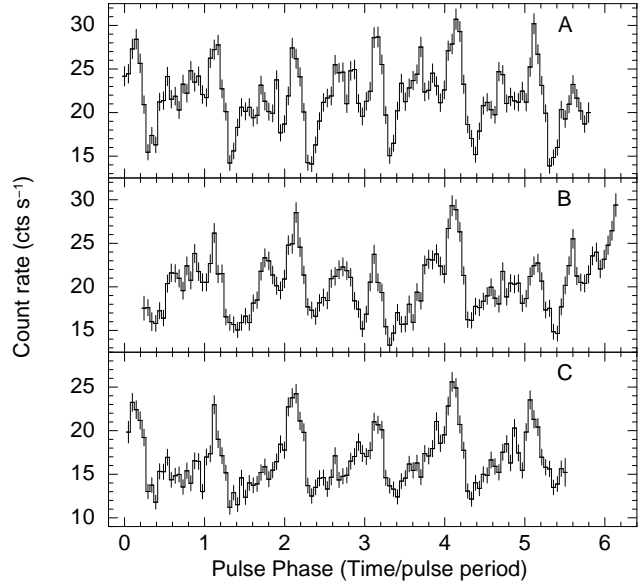}} 
  \caption{26 s binned light curves of three different \emph{RXTE}$-$PCA observations (3$-$20 keV, PCU2 top layer). Time 
  values are converted to phases (or time/pulse period) for arbitrary observation epochs. The IDs and mid-times of the 
  observations are A) 95044-05-02-02 at MJD 55404.6, B) 95044-05-02-03 at MJD 55406.7, C) 95044-05-03-00 at MJD 55408.5.}
  \label{lcex}
\end{figure}

A double-peaked pulse profile is observed in the first five \emph{RXTE}$-$PCA observations of SWIFT J1729.9$-$3437 
(see panel (A) of Fig. \ref{lcex}); however one peak loses its intensity starting from the middle of the observation on 
MJD 55406.7 (see panel (B) Fig. \ref{lcex}) as the source flux continues to decrease after the burst. The peak value of 
2$-$10 keV unabsorbed flux is $3.04 \times 10^{-10}\,$erg$\,$cm$^{-2}\,$s$^{-1}$ on MJD 55398.7, during its gradual decrease 
it reaches $1.96 \times 10^{-10}\,$erg$\,$cm$^{-2}\,$s$^{-1}$ on MJD 55408.5, when the shape of the pulse profile changes 
(see panel (C) Fig. \ref{lcex}). The minimum flux observed for the last \emph{RXTE}$-$PCA observation on MJD 55416.4 is 
$1.36 \times 10^{-10}\,$erg$\,$cm$^{-2}\,$s$^{-1}$. These flux values are calculated from the model flux of individual 
spectral fitting of each observation. 

\begin{figure*}
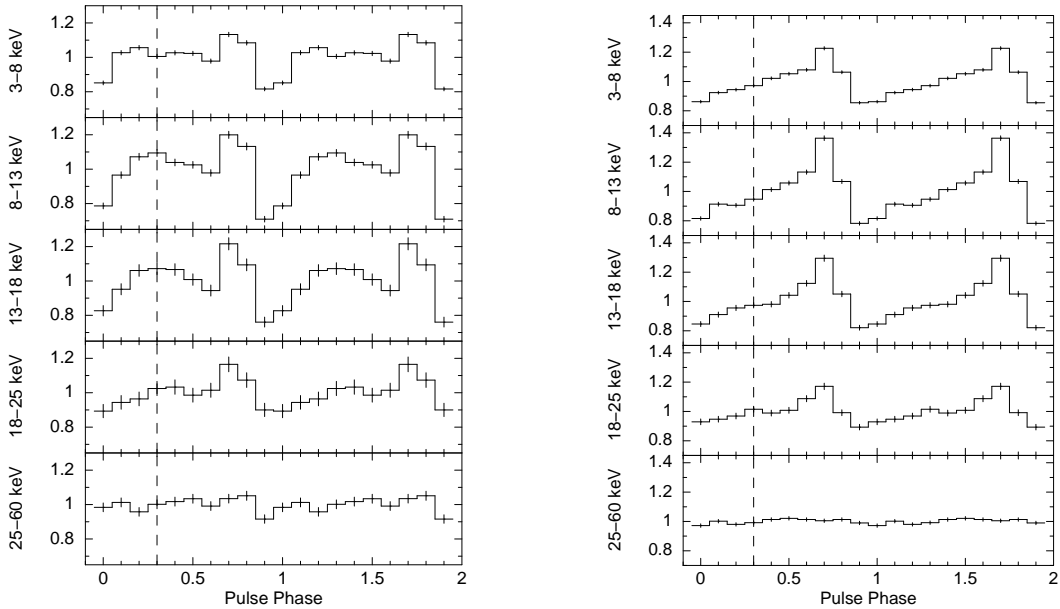

  \center{\includegraphics[width=8cm, angle=270]{3_norm_dash.eps}\hspace{1cm}\includegraphics[width=8cm, angle=270]{7_norm_dash.eps}} 
  \caption{Energy resolved pulse profiles of \emph{RXTE}$-$PCA observations on MJD 55400.78 (left; ID:95044-05-02-00) and 
  MJD 55408.46 (right; ID:95044-05-03-00). The unit of y-axis of each plot is normalized count from the specified energy 
  interval. The dashed line represents the mid-point of the secondary peak.}
  \label{enrespp}
\end{figure*}

To search for a possible energy dependence of the pulse shape change, we construct pulse profiles in five energy bands, 
i.e. 3$-$8, 8$-$13, 13$-$18, 18$-$25 and 25$-$60 keV. Two examples for the energy resolved pulse profiles, 
one for double-peaked and one for single-peaked, are given in Fig. \ref{enrespp}. We find that 8$-$13 keV and 13$-$18 keV 
pulses are stronger (they have higher pulse fraction) than the 3$-$8 keV and 18$-$25 keV pulses in all observations. In 
25$-$60 keV the pulse fraction noticeably drops. The pulse shape shows no strong energy dependence except for the two 
observations on MJD 55400.78 (ID:95044-05-02-00 see Fig. \ref{enrespp}) and MJD 55404.63 (ID:95044-05-02-02). In these two 
exceptions the secondary peak around pulse phase 0.3 loses its intensity at the 18$-$25 keV energy band.

\begin{figure*}
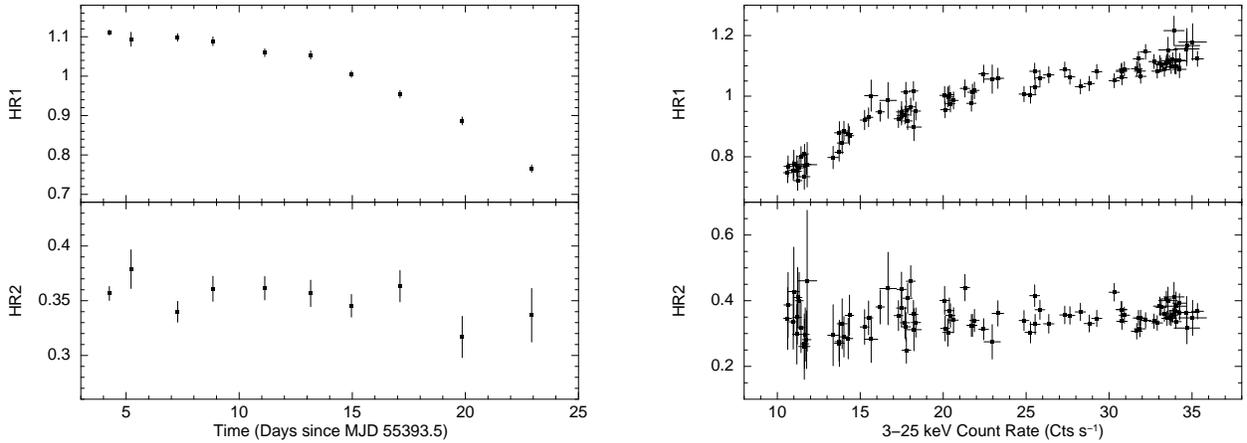

  \center{\includegraphics[width=5.8cm, angle=270]{hard_day.eps}\hspace{1cm}\includegraphics[width=5.8cm,
  angle=270]{hard_intensity.eps}} 
  \caption{Hardness ratios of energy resolved light curves. Daily averaged time evolutions are plotted on the left, whereas 
  530 s binned hardness ratios over 3-25 keV PCU2 count rates are plotted on the right. Specified abbreviations on y-axis 
  of each panel stand for HR1: 8$-$13 keV/3$-$8 keV, HR2: 18$-$25 keV/13$-$18 keV.}
  \label{hard}
\end{figure*}

We construct hardness ratio plots from the energy resolved light curves. Daily averaged count rates from 3$-$8, 
8$-$13, 13$-$18 and 18$-$25 keV light curves are used to plot hardness ratios HR1: 8$-$13 keV/3$-$8 keV and 
HR2: 18$-$25 keV/13$-$18 keV over time (see left panels of Fig. \ref{hard}). HR1 shows a noticeable evolution with respect 
to time, it remains almost constant at a value $\sim1.1$ until MJD 55406.7 and a gradual decrease starts after the sixth 
observation. The interval with constant HR1 correspond to times when the pulse profile is double-peaked, where as the 
decreasing interval of HR1 coincides with the times when the pulse profile is single-peaked. Nevertheless, it should be 
noted that the pulse profiles show no significant variation in the corresponding energy bands (see Fig. \ref{enrespp}). 
Furthermore, 530 s binned hardness ratios are plotted over the total count rate in 3$-$25 keV band (see right panels of 
Fig. \ref{hard}). A noticeable correlation is again observed for HR1. As the source flux decreases during the ongoing decline 
of the outburst, the emitted radiation becomes softer.

\section{Spectral Analysis}

\subsection{Overall Spectrum}

\begin{figure}
  \center{\includegraphics[width=6.2cm, angle=270]{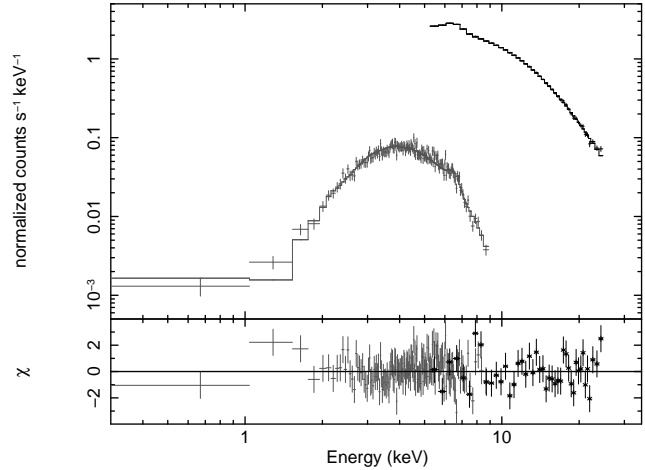}} 
  \caption{Simultaneous spectral fitting of \emph{Swift}$-$XRT (0.3$-$9.3 keV; grey) and \emph{RXTE}$-$PCA (5$-$25 keV; 
  black) data. The data and its best fit with \texttt{WABS$\times$(CUTOFFPL$+$GAU)} ($\chi^{2}=1.09\,$; solid line) are 
  shown in upper panel, the residuals are given in the lower panel.}
  \label{bothspe}
\end{figure}

\begin{table}
  \caption{Best fit (\texttt{WABS$\times$(CUTOFFPL$+$GAU)}) spectral parameters for the simultaneous fitting of 
  \emph{Swift}$-$XRT and \emph{RXTE}$-$PCA data shown in Fig.\ref{bothspe}. $C_1$ and $C_2$ are the multiplicative constant
  factors for different instruments. All uncertainties are calculated at the 90 per cent confidence level. }
  \label{simspe}
  \center{\renewcommand{\arraystretch}{1.5}\begin{tabular}{llr}
  \hline \hline	
\bf{Parameter} 		&	& \bf{Value} \\	
 \hline
 $C_1$   	& for XRT				 & 1.00 (fixed) \\
 $C_2$ 	  	& for PCA				 & $1.55^{+0.04}_{-0.04}$ \\
 $n_H$ 		& $(10^{22}\,$cm$^{-2})$ 		 & $8.27^{+0.37}_{-0.36}$ \\
 $\Gamma$   	&					 & $1.23^{+0.07}_{-0.07}$ \\
 $\Gamma$ Norm. &($10^{-2}\,$cts$\,$cm$^{-2}\,$s$^{-1}$) & $2.42^{+0.26}_{-0.25}$ \\
 $E_{fold}$ 	&(keV)  				 & $16.50^{+1.79}_{-1.52}$ \\
 $E_{Fe}$ 	&(keV)   				 & $6.51^{+0.03}_{-0.03}$ \\
 $\sigma_{Fe}$ 	&(keV)					 & $0.25^{+0.06}_{-0.05}$ \\
 $Fe$ Norm. 	&($10^{-4}\,$cts$\,$cm$^{-2}\,$s$^{-1}$) & $4.68^{+0.49}_{-0.41}$ \\
 Unabs. $F\,_{0.3-9.3\text{ keV}}$ & $(10^{-10}\,$erg$\,$cm$^{-2}\,$s$^{-1})$ & $2.08^{+0.017}_{-0.024}$ \\
 Unabs. $F\,_{5-25\text{ keV}}	 $ & $(10^{-10}\,$erg$\,$cm$^{-2}\,$s$^{-1})$ & $3.00^{+0.004}_{-0.016}$ \\
\hline
Reduced $\chi^{2}$ (dof) &	& 1.09 (187) \\
\hline
\end{tabular}} \\
\end{table} 

A preliminary spectral analysis for the first pointed observations of SWIFT J1729.9$-$3437 was performed by Markwardt 
et al. (2010b). In this paper, we extend the spectral study by using all the available observations of the source defined 
in Section 2. Basically, the spectra can be modelled by a power law with a high energy cut-off and photoelectric absorption 
as it is suggested by Markwardt et al. (2010b). Among the several models that describe the cut-off power law, the best fit 
is achieved by \verb"CUTOFFPL" model in \verb"XSPEC". An additional Gaussian component is also required for a weak \emph{Fe} 
emission line around 6.4 keV. During the simultaneous fitting of \emph{Swift}$-$XRT and \emph{RXTE}$-$PCA spectra we included 
a multiplicative constant factor in the model to account for the normalization uncertainty between the two instruments. The 
data and its best fit are plotted in Fig. \ref{bothspe} and the corresponding spectral parameters are given in 
Table \ref{simspe}. 

The energy ranges for the simultaneous spectral analysis are initially selected to be 0.3$-$9.3 keV for the XRT spectrum and 
3$-$25 keV for the PCA spectrum. Individually these spectra have similar shapes that can be modelled with the same models, 
apart from an offset in absolute flux calibration. However individual modelling of the data from different instruments yields 
different absorption parameters due to different instrumental band-passes (Markwardt et al. 2010b). During the simultaneous 
fitting trials we observe large residuals for the first energy bins of the PCA spectrum although the fit is adequate for the 
XRT spectrum. Therefore we exclude energies below 5 keV for the PCA spectrum, since the XRT spectrum has more spectral 
energy bins in soft X-rays. 

The FOV of PCA is large and SWIFT J1729.9$-$3437 is near the Galactic ridge. The thermal emission from the ridge is usually 
known to contaminate the spectral count rates of the source. During the analysis of the first PCA observation, the weak line 
at $\sim$6.6 keV reported by Markwardt et al. (2010b) was suggested to be a contamination, since it could not be resolved in 
the spectrum of the first XRT observation. We tried to handle this issue in overall PCA spectrum with fixed additive 
Galactic ridge models. We confirm that the flux of the source is two orders of magnitude bigger than the flux of the 
Galactic ridge (Valinia \& Marshall 1998) therefore we conclude that the contamination could be small enough to be neglected. 
Furthermore; addition of the Gaussian model component for the \emph{Fe} line improves the individual fit of the 0.3$-$9.3 keV 
XRT spectrum after combining data from all XRT observations by reducing the reduced $\chi^{2}$ from 1.08 
($\chi^{2}/$d.o.f. $ = 159.0 / 147$) to 0.98 ($\chi^{2}/$d.o.f. $ = 143.8 / 146$). The F-test probability that this 
improvement is achieved just by chance is $1.3 \times 10^{-4}$. Therefore we suggest that the \emph{Fe} emission originates 
from SWIFT J1729.9$-$3437. The line energy found for the source is shifted from the neutral value of \emph{Fe} K$\alpha$ 
($\sim$6.4 keV), which may be a consequence of an excess of ionization.

\begin{figure}
  \center{\includegraphics[width=6.2cm, angle=270]{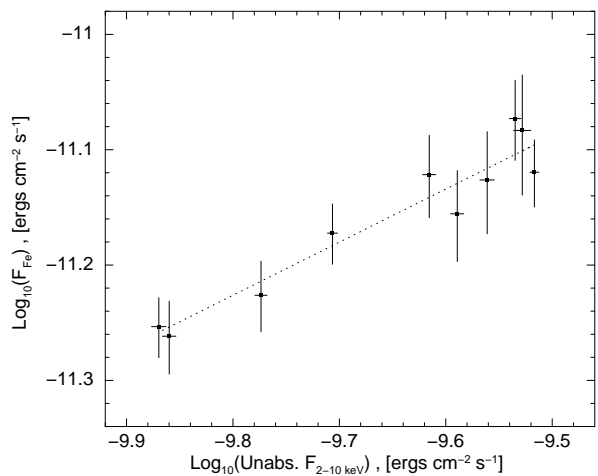}} 
  \caption{Variation of \emph{Fe} line flux with the X-ray flux from \emph{RXTE}$-$PCA observations. The error bars indicate 
  uncertainties at 1$\sigma$ (68 per cent) confidence level. The dotted line represents the linear fits to the data.}
  \label{Fe}
\end{figure}

In accretion powered pulsars, the \emph{Fe} K$\alpha$ line is produced by the reprocessing of the hard X-ray emission in the 
relatively low ionized and cool matter surrounding the pulsar. The correlation of the iron line intensity and the continuum 
intensity in many sources (e.g. Vela X$-$1: Ohashi et al 1984, GX 301$-$2: Leahy et al. 1989) is specified to be the evidence 
for the fluorescence of X-ray continuum by the cold material. As the continuum intensity increases the illumination of the 
matter increases and the line intensity strengthens. We measure \emph{Fe} line flux values from the individual \emph{RXTE} 
observations of SWIFT J1729.9$-$3437 by using the \verb"CFLUX" model in \verb"XSPEC" and find that they correlate with the 
source flux values (see Fig. \ref{Fe}). In Fig. \ref{Fe} the dotted line represents the linear fits to the data, with a 
slope of $0.46\pm0.12$. The Pearson product-moment correlation coefficient between \emph{Fe} line flux and unabsorbed source 
flux at 2$-$10 keV is 0.96. The null-hypothesis probability calculated from the Student's t-distribution (two-tailed) is 
$1.2 \times 10^{-5}$. This correlation also confirms the origin of the \emph{Fe} emission.

\subsection{Pulse Phase Resolved Spectra}

We construct pulse phase resolved spectra of SWIFT J1729.9$-$3437 from the first three RXTE$-$PCA observations. There are 
two motivations to the data selection. First, the brightest of the observations are selected to ensure the best 
signal-to-noise ratio possible. Second, the selected observations are the only ones that have 256 channels Good Xenon event 
mode data available, which is required for the \verb"FASEBIN" tool. The timing solution found for the source in Section 3.1 
is appended to the timing files of the \verb"FASEBIN" tool for correct phase-binning. The pulse period is divided into five 
equal segments resulting in $\sim$2.1 ks exposure time for each phase bin spectrum.

\begin{figure}
  \center{\includegraphics[width=10.2cm, angle=270]{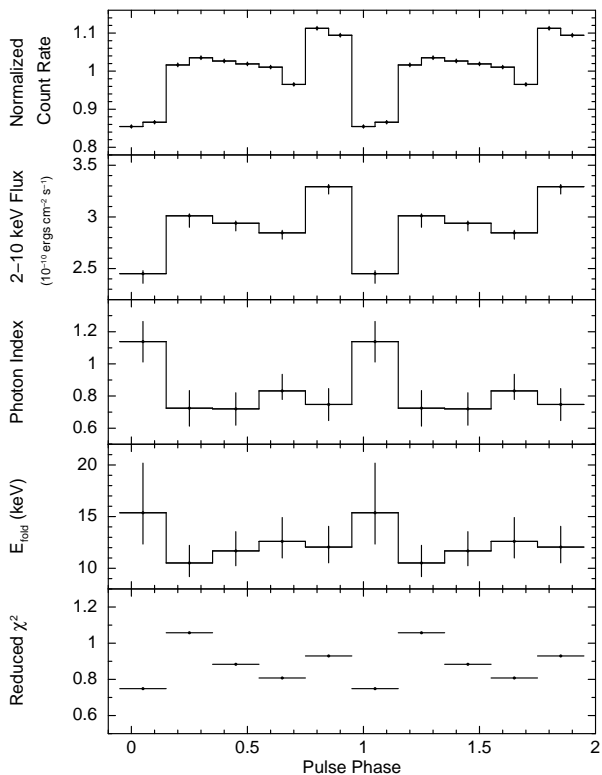}} 
  \caption{Pulse phase resolved variations of the spectral parameters with 5 phase bins for the first three
  \emph{RXTE}$-$PCA observations. The pulse profile is plotted in 10 phase bins at the top panel. From top to bottom, we show 
  the pulse profile, 2$-$10 keV unabsorbed flux, the photon index, the folding energy of exponential roll-off and reduced 
  $\chi^{2}$ of the fits. All uncertainties are calculated at the 90 per cent confidence level. For clarity, the data points 
  are repeated for a cycle.}
  \label{phasespe}
\end{figure}

We model 3$-$25 keV spectra with the same models used for the overall spectrum in the previous section. As we do not observe 
any significant change in the Hydrogen column density ($n_H$) and the \emph{Fe} line energy during the preliminary fitting 
trials, we fixed these parameters during the analysis in order to better constrain the other parameters of the fits. The 
photon index and the folding energy of the high energy cut-off ($E_{fold}$) are found to be varying within pulse phases 
(see Fig. \ref{phasespe}). The phase dependence of photon index is in anti-correlation with the pulsed flux. The Pearson 
product-moment correlation coefficient between pulsed flux and photon index values is $-0.87$ and the corresponding 
null-hypothesis probability (two-tailed) is 0.06. The null-hypothesis probability of this anti-correlation is quite 
significant but our results still indicate that the softest emission is observed at the un-pulsed phase around 0.05; from 
which we might suggest a possible physical relation between the parameters. The $E_{fold}$ values show a similar trend as 
the photon index values, which means softer spectra have higher cut-off energies. The Pearson product-moment correlation 
coefficient between $E_{fold}$ and photon index values is 0.96 and the corresponding null-hypothesis probability (two-tailed) 
is 0.01. Although the correlation analysis indicates a strong dependence between the parameters, one should note that the 
uncertainties in the parameters are not taken into account during this analysis. When the large uncertainties in $E_{fold}$ 
values are taken into account the data is consistent with a constant value of $\sim11.7$ keV. Therefore it is difficult to 
infer a clear variation of the $E_{fold}$ with the pulse phase, suggesting a rather marginal detection.

\section{Summary and Discussion}

In this paper, we study timing and spectral properties of SWIFT J1729.9$-$3437 using \emph{RXTE} and \emph{Swift} 
observations of its outburst from 2010 July 20 through 2010 August 12 (MJD 55397 $-$ MJD 55420). From these observations 
with a time span of $\simeq 23$ days, we find that arrival times can be fit to a quadratic. From this fit, we calculate a 
spin frequency and spin frequency derivative of $1.8734(8) \times 10^{-3}$ Hz and $6.42(6) \times 10^{-12}$ Hz/s 
respectively. The residuals of the quadratic is further found to fit a circular orbital model with 
${{a} \over {c}}\sin i=65(3)$ lt.s and an orbital period of 15.3(2) days. We also try an elliptical model and find an upper 
limit for eccentricity as 0.60. However, this model over-fits the data with a reduced $\chi^2$ of 0.40. Future observations 
might help to refine orbital parameters of the source.

Using this ${{a} \over {c}}\sin i$ value, we find the mass function 
(${{4\pi^2} \over {G}}{{(a\sin i)^3} \over {P_{orbital}^2}}$) to be about $1.3M_{\odot}$. An orbital period of 15.3 days and 
a spin period of 533.76 s puts the source in line with the accretion powered pulsars with supergiant companions, in the
Corbet diagram (Drave et al. 2012, Corbet 1984). On the other hand, the small mass function obtained from the circular 
orbital model should be an indication of a small orbital inclination angle. If the circular orbital model is preferred as 
the model fitting the residuals of the quadratic fit, this indicates that we observe the binary system nearly edge-on. 

Alternatively the residuals of the quadratic fit can be explained by a torque noise strength of 
$6.8 \times 10^{-18}$ Hz sec$^{-2}$. This value is quite consistent with other accreting X-ray binaries 
(Baykal \& \"{O}gelman 1993, Bildsten et al. 1997). Future observations are needed to understand the exact nature of the 
source.

Initially, a double-peaked pulse profile is observed in the light curves of the source. Then we find that one peak loses its 
intensity starting from the middle of the observation on MJD 55406.7 as the source flux continues to decrease after the 
burst. To study the energy dependence and temporal variability of the pulse profiles, we construct pulse profiles with five 
different energy bands shown in Fig. \ref{enrespp}. We observe stronger pulses in the 8$-$13 keV and 13$-$18 keV energy bands
but generally the pulse shape shows no strong energy dependence. Double to single peak transition seen in this source might 
be due to a sharp decline in the intensity of the radiation coming from the fan beam, since the formation of fan beam 
strongly depends on the luminosity of the source as for EXO 2030+375 (Parmar et al. 1989) and GX 1+4 (Paul et al. 1997). 
However lack of significant energy dependence of the pulse profiles makes the fan beam explanation implausible since fan 
beams are expected to be spectrally harder than pencil beams.

In order to have a basic understanding of spectral variability of the whole observations, we study hardness ratios of the 
source. From the hardness ratio plots (see Fig. \ref{hard}), we suggest that the emitted radiation becomes softer as the 
source flux decreases during the ongoing decline of the outburst. The similar spectral softening with decreasing flux was 
reported before for 1A 0535+262, A 1118$-$616, SWIFT J1626.6$-$5156, XTE J0658$-$073 and GRO J1008$-$57 using 
hardness-intensity relations (Reig \& Nespoli, 2013). 

To extend the preliminary spectral analysis of Markwardt et al. (2010b), we also construct a single spectrum from all the 
available \emph{RXTE} and \emph{Swift} observations (see Fig. \ref{bothspe} and Table \ref{simspe}). We find that adding an 
\emph{Fe} line complex feature with a peak at 6.51 keV slightly improves the spectral fit. We discuss that this \emph{Fe} 
line feature is more likely originated from the source as the galactic ridge emission is too weak to explain this emission 
alone. We also measure \emph{Fe} line flux values from the individual \emph{RXTE} observations of the source and find that 
they correlate with the source flux values (see Fig. \ref{Fe}). This correlation confirms the origin of the \emph{Fe} 
emission as the source itself rather than the galactic ridge.

We perform pulse phase resolved spectral analysis of the source using the first three \emph{RXTE} observations. From this 
analysis, we find a marginal evidence of a variation of the photon index and the folding energy of the high energy cut-off 
with the pulse phase (see Fig. \ref{phasespe}).

\section*{Acknowledgements}

We acknowledge support from T\"{U}B\.{I}TAK, the Scientific and Technological Research Council of Turkey through the 
research project TBAG 109T748.

\bsp

\label{lastpage}

\end{document}